\documentstyle[floats,prb,epsf,aps,twocolumn]{revtex}

\newcommand{\ba}{\begin{eqnarray}}
\newcommand{\ea}{\end{eqnarray}}
\newcommand{\be}{\begin{equation}}
\newcommand{\ee}{\end{equation}}

\begin{document}
\wideabs{
\draft
\title{Quantum Hall Effect: Current Distribution and Existence of
Extended States}

\author{K. Tsemekhman, V. Tsemekhman}
\address{Department of Physics, University of Washington,
 Box 351560, Seattle, Washington 98195}
\author{C. Wexler}
\address{Department of Physics, Box 118440, University of Florida,
 Gainesville, Florida 32611}
\date{December 1998}

\maketitle 
\begin{abstract} 
We present a consistent description of the current distribution in the
quantum Hall effect, based on two main ingredients: the location of the
extended states and the distribution of the electric field.  We show that
the interaction between electrons produces a boundary line below the Fermi
energy, which extends from source to drain.  The existence of this line
and that of a physical boundary are responsible for the formation of a
{\em band} of extended states that carry the Hall current. The number and
density of these extended states are determined by the difference between
the energy of this equipotential boundary line and the energy of the
single extended state that would exist in an infinite system. This is used
to prove that the band of extended states is distributed through the bulk
of the sample. We explore the distribution of the Hall currents and
electric fields in by presenting a model that captures the main features
of the charge relaxation processes. Theoretical predictions based on this
model and on the preceding theory are used to unambiguously explain recent
experimental findings. 
\end{abstract} 
} 

\section{Introduction}
\label{Intro-sex}

Since the discovery of the quantum Hall effect (QHE)~\cite{klitz} two
different approaches have been used to describe it: one focuses on the
properties of the {\em edge channels} \cite{Halperin,buttiker}, while 
the other concentrates on the states in the {\em bulk} of the system
\cite{aoki,avron,Thouless,wexler,ando,ruzin}.
Both of these approaches associate the quantization of the transverse
conductivity $\sigma_{xy}$ in units of $e^2/h$, and the absence of
dissipation ($\sigma_{xx}=0$) with the formation of an incompressible
liquid state in the bulk of the system, and both are based on the
existence of delocalized states. 
However, the nature of the injected Hall current
is quite different in these theories. 
In the first approach
\cite{Halperin,buttiker}, which neglects inter-electron interaction,
electrons from the current source are injected into delocalized states
{\it at the Fermi level} which carry the Hall current. Since
the only states at the Fermi level are located at the edge of the
system (edge channels), the Hall current has to be confined to this
narrow region near the sample boundary. The magnitude of this edge
current is $ne^2/h$ times the difference between the chemical
potentials at the two edges of the system. 
In the latter approach the Hall current density is
proportional (with the same constant of proportionality) to the 
electric field, whose appearance is due to the charging of the
system \cite{aoki,avron,Thouless,wexler,ando,ruzin}. In this case
the Hall current is still the same but it generally flows both along
the edge and in the bulk of the system depending on the Hall electric
field distribution. 

Experimentally, it has proved difficult to
determine which of these descriptions is correct, and different
probes produce apparent contradictory answers.
Measurements of the equilibration rates between the
current carrying states~\cite{alphenaar}, as well as experiments on
non-local conductivity~\cite{Goldman} have natural explanations in
terms of edge currents. However, studies of the breakdown of the
QHE~\cite{israel}, the direct measurements of the Hall voltage in the
sample~\cite{Tsui} and recent measurement of electric field
distribution \cite{Ashoori,MacEuen,Yacobi} favor the picture of bulk
currents.
It is the objective of this paper to clarify the situation and provide
a description of the current distribution in the quantum Hall regime
which is consistent with the seemingly opposing experimental results. 

It was proved earlier \cite{We_SSC} that the Hall current is absent in
the regions near the edge where the electron density increases from
zero to its bulk value. These are the conventionally defined edge
channels. Yet one can imagine the situation
when most or all of the current does flow right next to these regions,
along their boundaries. 
Moreover, the existence of the disorder potential
modifies the shape and the whole notion of edge channels as they
can no longer be thought of as simple homogeneous strips `parallel' to
the boundaries. 

The complexity of the near-edge as well as of the bulk
geometric structure was revealed by recent experiments on imaging the
potential distribution in the QH bar at different filling factors
\cite{Ashoori,MacEuen,Yacobi}. 
It is worth noting that the relation between the Hall electric field
and Hall current in such a complex inhomogeneous system can only be
reasonable with some kind of averaging procedure. Local currents are
the combination of non-equilibrium Hall current and diamagnetic
currents and, in general, it is not possible to differentiate them
without averaging over distances of the order of a typical 
localization length of the closed trajectories. A similar argument
holds for the disorder and Hall voltage contributions to the electric
potential. Therefore one has to be very careful inferring the Hall
current distribution from the measured electrostatic potential
distribution. When the Hall voltage is sufficiently large the external
electric field smears out some of the disorder potential fluctuations
and observation of this electric field distribution on shorter scales
becomes possible. However, the state of the QH system in such
strong Hall electric field is far from equilibrium, and the information
obtained from non-linear regime measurements, although interesting and
valuable in itself, does not interpolate into a linear regime
(experiments associated with the study of the breakdown of the IQHE
produced current-voltage characteristics very distinct from linear at
sufficiently high Hall voltages \cite{komi,Nachtwei}).

The problem of the current distribution in the QH sample is also
intrinsically connected with the question of the existence of
delocalized states in the presence of disorder. The latter was widely
discussed in the literature
\cite{Lurye,Trugman,Khmelnitskii,Laughlin,Huckenstein} 
but was not usually related to the problem of current distribution.  
Trugman \cite{Trugman}  showed the connection between the existence
of delocalized states  and the classical percolation problem in an
infinite system with long-range disorder potential. In his paper
\cite{Trugman} and in Ref.\ \onlinecite{Lurye} 
the possibility of creation of current carrying extended
states by external electric field is discussed. The role of the physical
boundaries of the system on the existence of delocalized
states was also addressed in Ref.\ \onlinecite{Lurye}. However, it was
still not clear where these delocalized states 
are located with respect to the physical boundaries of the system, and
whether the delocalization due to the applied external electric field
is confined to the edges or not. 

In this paper we examine the distribution of extended trajectories
in equilibrium and use these findings to study the distribution of the
Hall current in the bulk. The conditions for the electric field to be
concentrated at the boundaries or spread into the sample are found and
compared with corresponding experimental values.
We find that, in addition to the restrictions put forth in 
our previous work \cite{We_SSC}, the Hall current paths are distributed
throughout the bulk of the system under the most common experimental
conditions. 
The interplay between the inter-electron interaction and the long-range
disorder potential turn out to be crucial in establishing such a
current distribution. 

The paper is organized as follows. In Sections \ref{sec:Equil} and
\ref{sec:Geom} we discuss the equilibrium state of a Quantum Hall
system when no Hall current is present. We prove that there are
delocalized states in  incompressible regions both near the edges and
in the bulk of the sample. 
Section \ref{sec:steady} deals with the Hall field and Hall current
distributions away from equilibrium. We find that, similarly to
the situation in equilibrium, extended states are distributed
throughout the sample, their particular spatial density being
dependent on disorder configuration and on the rate of relaxation
processes.
In Section \ref{sec:Streda} we discuss the
validity of {St\v{r}eda} formula \cite{Streda} in the context of previous
arguments. The controversy around the formula is resolved using the
arguments of the previous sections.
Finally, in Section \ref{sec:Expts} a consistent interpretation of
recent experiments is presented. Although the findings of these
experiments do not appear to be in complete agreement with the absence
of near-edge Hall current, we demonstrate how these results confirm the
theory of the distribution of extended states presented in this 
paper when all the experimental conditions are fully accounted for.

\section{Compressible and Incompressible Bulk}
\label{sec:Equil}
 
Consider a typical system where the QHE is observed: a
2DEG formed at the interface of the GaAs/GaAlAs heterostructure,
placed in a strong magnetic field $B$, with the length of the system
$L_{y}$ being much larger than its width $L_{x}$. In absence of
disorder, in the Landau gauge, with vector potential $A_{y}=B\,x$, the
electrons are free in the $y$-direction and quantized in the 
$x$-direction.  Wave functions can be written as
$e^{iky}f_{n}(x-kl^2)$,  where $l=(\hbar c/eB)^{1/2}$ is the magnetic
length and $n$ enumerates the Landau levels (LL's). In homogeneous
systems, wave functions and energies are simply those of the harmonic
oscillator: $\epsilon_{n}=\hbar \omega_{c}(n+1/2)$, where
$\omega_{c}=eB/mc$ is the cyclotron frequency.

The disorder potential plays a crucial role
in establishing the QHE. Two different types of disorder potential,
short-range and long-range, require different approaches to elucidate 
the properties
of the 2DEG. The former is realized, e.g. in the MOSFET-devices
and InAs heterostructures; it leads to low-mobility samples and is
known to produce wide QH plateaus and scaling in the transitional
regimes between the QH plateaus even at temperatures above 1~K
~\cite{Mosfet}. High-mobility samples are formed at the interface
of GaAs--Al$_{x}$Ga$_{1-x}$As heterostructures, characterized by a
long-range disorder potential. This potential is created by a
Si-ion layer separated from the 2DEG by a undoped spacer with a
typical thickness $d\simeq 100 - 500 $\AA  
~\cite{Prange1}. 
In this paper we will be concerned only with this
latter type of the systems. In contrast to 
low-mobility samples, the inter-electron interaction is very important in
the case of a long-range disorder potential: electrons tend to
screen potential fluctuations whose wave-length $\lambda$
is larger than both the magnetic length (see Refs.\
\onlinecite{Efros,We_DOS,Dima-Th}).  At the same time the finite distance to
the donor plane strongly suppresses fluctuations with wavelengths
shorter than the spacer thickness~\cite{Efros,We_DOS}.

In presence of a strong magnetic field, however, screening is
limited by the incompressibility associated with the existence of the
gap between the LL's, and filled LL's do not contribute to screening.
Still, if the filling factor $\nu=\pi l^2 n_0$, $n_0$ being the
electron density, is not close to an integer, there are plenty of
unoccupied states on the top-most occupied Landau level, and the
long-range fluctuations of the random potential get screened in a
significant part of the sample area \cite{Efros}. When the filling
factor is close to an integer one expects to find regions in which
the fluctuations of the random potential are not screened. In these
regions some LL's are completely filled while the rest are completely
empty. When a region with $N$ completely filled LL's percolates
throughout the sample the QHE is observed. The number $N$ is usually
the integer closest to the average filling factor $\nu$. 

Because this percolation picture of the QHE is valid only for filling
factors close to an integer, it cannot describe the whole range of
magnetic fields where the QHE is observed. Evidently, some other
aspects of this problem, not accounted for in this simple picture,
lead to the QHE at  occupation numbers away from integers.  
This important issue will be discussed in an upcoming publication
\cite{We_DOS}. 

In this work, we restrict ourselves to systems in which the region
occupied by the incompressible liquid percolates, and study the
current distribution in this percolating region. The potential at
every point of the sample is determined by the random distribution of
impurities in the dopant layer and by the distribution of the electron
charge all over the 2DEG-plane. The potential thus 
generated has only relatively
long-wavelength variations, and this, combined
with the strong magnetic field, allows for the following simplification:
electrons can be considered drifting along the lines of the constant
classical energy \cite{Prange}. 
The Hall current injected into the sample is therefore
carried by the electrons moving along the trajectories
stretching from source to drain. Thus,
there are two factors that eventually determine the spatial
distribution of current in the sample: the steady state charge
and Hall electric field distributions, and the location of 
the extended ({\it i.e. percolating from source to drain}, the
definition which we will be using throughout the paper) classical
trajectories. To study these properties in the steady 
state, we first address the same questions for the
equilibrium state of this 2DEG system.

\section{Extended trajectories and diamagnetic current distribution in
equilibrium}
\label{sec:Geom}

In this section we develop a consistent description of the current
carrying states in equilibrium. It is proven that extended
states are present both deep in the bulk of the sample and 
close to its edges.

\subsection{Edge channels in presence of disorder: definition 
and absence of Hall current}
\label{subsec:start}

\begin{figure}
  \begin{center}
    \leavevmode
    \epsfxsize=4cm 
    \epsfbox{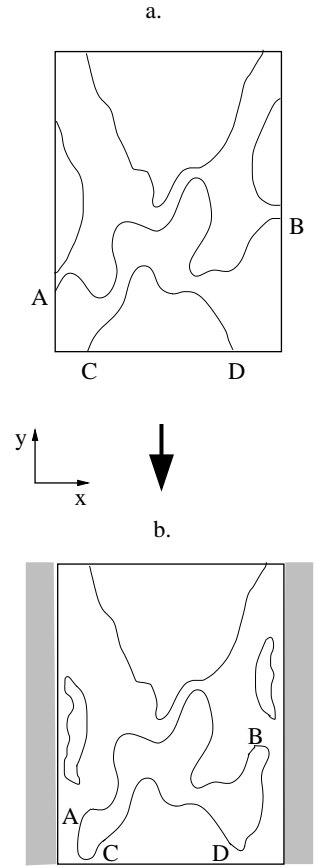}
  \end{center}
       \caption{Finite size effects on the extended states.
                 a. Elimination of percolation in the longer
                 $y$-direction. Trajectory $AB$ is one from the 
                 band of the states percolating in x-direction;
                 line $CD$ is part of a closed trajectory in
                 the corresponding infinite system.
                  b. Effects of a confining potential: lines $AB$ and
                  $CD$ link to each other to become a single closed
                  equipotential line. Other trajectories are also
                  deformed by the confining potential.
                 {\label{bound_eff}}}
\end{figure}

Let us consider an infinite two-dimensional random potential. It is
clear from percolation theory that all equipotential lines but one are
closed~\cite{Lurye}. As a corollary we find that in an
{\em infinite} 2DEG in a strong magnetic field for each LL there is
a single energy $E_{i}$ such that the state with this energy is extended in
both directions.  The only percolating equipotential line corresponds to
this extended state (one per each LL). If a finite-size system were
equivalent to a  piece of the same size in the infinite system (with
no inhomogeneities introduced by the boundaries), a band of states
around $E_{i}$ whose localization length is larger than the sample
size, would appear as delocalized. However, in the finite size sample
these states can percolate only in one direction, namely, the shorter
one~\cite{Lurye}. The simple assumption above produces a drastic
consequence: there are no extended states in the longer $y$-direction,
and we should expect no Hall current in this direction (Fig.\
\ref{bound_eff}a). It is clear that the boundaries produce a
non-trivial effect.


\begin{figure}
  \begin{center}
    \leavevmode
    \epsfxsize=8cm
    \epsfbox{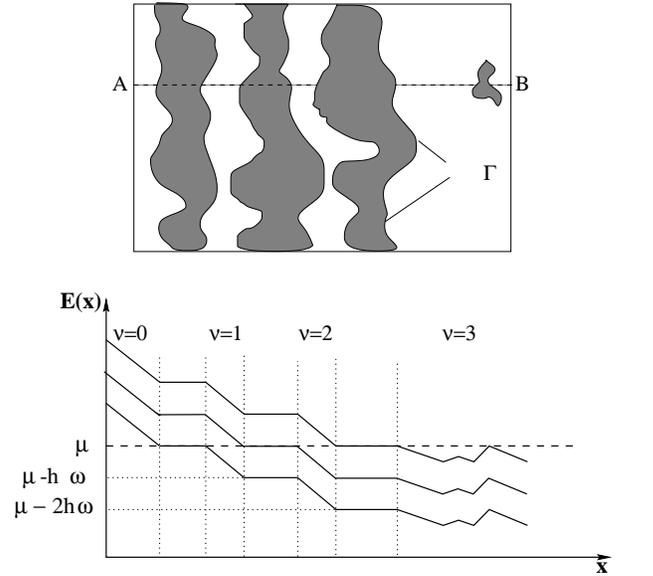}
  \end{center}
    \caption{ \label{near_edg}
       Charge distribution and energies of Landau levels near
       the physical boundary
       a. Compressible and incompressible `strips' near the edge.
       b. Energies of the first three occupied Landau levels close to
          the sample boundary. }
\end{figure}

In reality, the 2DEG is confined in the shorter $x$-direction by a
self-consistent potential created by the positive background, electron
charge distribution, and gate voltages. Such a potential leads to an
increase of the density of electrons from zero at the physical edge of
the 2DEG to the bulk value. Depending on the sharpness of the
confining potential such growth can be either gradual or step-like 
\cite{Chklovskii,Wen_Chamon}. For illustrative purposes, let  
us assume that the charge density grows gradually (but the arguments
that follow are valid in either situation). 
In this case, a narrow compressible region with a partially filled LL
percolating in $y$-direction is formed
near the edge~\cite{Chklovskii}. In this region the
variations of the potential due to confinement and disorder are largely
screened by electrons. As long as there are enough
unfilled states on the same LL, electrons can screen variations 
of the potential. The density profile $n(x)$ in the 2DEG at the edge
is approximately given by the electrostatic solution \cite{Chklovskii}
\be
  n(x) = \left [ \frac{x - l}{x + l} \right ]^{1/2} n_{0} \, , \,\,\,
     x > l.
\label{Chk_Den}
\ee
Here $x$ is the distance from the physical boundary of the 2DEG and
$l$ is the width of the depletion layer.
However, as the occupation number of the
first filled LL grows as required by the electrostatic
solution (\ref{Chk_Den}), less and less `space' is available 
for variations of the local electron density, and consequently, more
and more variations of the potential are left unscreened. Finally, as
we move towards the bulk of the system, the electrons completely loose
the ability to screen the potential fluctuations, and these unscreened
regions link to form a continuous strip of incompressible liquid along
the $y$-direction. The outermost incompressible strip corresponds to a
single filled LL. Moving further away from the edge, the potential
starts falling. 
If more than one LL are filled in the bulk, the potential drops by 
$ \hbar \omega_{c} $ in each incompressible strip, and then
another compressible river (with partial occupation in the next LL),
is formed roughly parallel to the $y$-direction~\cite{Chklovskii}.  
When the last ($N$-th) LL becomes completely filled, the fall of
potential is entirely due to the unscreened fluctuations of the
disorder potential (Fig.\ \ref{near_edg}b). Note that while a
purely electrostatic solution in the absence of disorder predicts a
complete screening in the bulk for any filling factor, the situation
is, in fact, considerably different due to the interplay of
disorder and LL quantization. 


We conclude that at the left edge of the sample there is a geometric
boundary line $\Gamma$ which ought to be infinite in the
$y$-direction, and which marks the transition from the last
compressible strip to the incompressible bulk (Fig.\ 2a). 
To the left of this line the density of the electron charge 
falls off, and there are $N$ percolating compressible strips. We will
refer to this region from the physical edge of the system to the
boundary line $\Gamma$ as to the `edge channel region' (there are two
such regions -- one at each edge as the argument above clearly holds for
the right edge, as well). On the other side of this boundary, an
incompressible region with $N$ completely filled LL's percolates. In
this region the potential falls due to the unscreened fluctuations of
the disorder potential (Fig.\ 2b). To this whole region we will refer
to as the bulk. 

Let us first show that in the regime of the QHE no part of the
non-equilibrium Hall current flows in the edge channel region as
defined above (for a more detailed discussion see Ref.\
\onlinecite{We_SSC}). 

In equilibrium, although the net current is zero, orbital current flows
along the trajectories extended from source to drain at
one edge and in the opposite direction at the other edge. 
There are two contributions to this
equilibrium current which are spatially separated \cite{Vignale}.
First, the diamagnetic current, is roughly proportional to the
gradient of the self-consistent confining potential, and it is non-zero
only in the incompressible regions. The other one is associated with the
concentration gradient, which is different from zero only in the
compressible strips close to the device boundary. However, the total
current carried by the {\it partially} occupied near-edge region of
the $n$-th LL is always $(n+1/2)e\omega_{c}/2\pi$, independent of the
external conditions \cite{Vignale}. Corresponding compressible strips
at opposite edges of the sample carry the same current but in opposite
directions,  therefore only incompressible regions can support
non-equilibrium current. This can also be seen from classical
considerations: non-equilibrium Hall current is 
carried by electrons drifting in crossed electric and magnetic fields
in the direction perpendicular to both of them. This drift velocity is
proportional to the local electric field, which is identically zero in the
compressible regions. We will, therefore, ignore the current in the 
compressible regions (even if extended trajectories are present there)
since it is not changed by the injected {\em non-equilibrium} current. 

Consider now the case when current is injected into the sample. Under 
conditions of the QHE, the longitudinal conductivity is negligible,
and there is no dissipative current between the two edges. Therefore,
for every value of the injected current smaller than the critical current
corresponding to the QHE breakdown, there exists a many-particle
{\em steady state} that carries the current. This non-equilibrium
steady state can be described by certain charge and current
distributions in ways similar to an equilibrium state. The only
difference is that in the steady state the net current in the
transverse direction is different from zero. As a 
consequence of this, LL's are filled up to different energies at
the left and the right edges, and a Hall voltage can be measured. As
the non-equilibrium current is carried by the states in the
incompressible regions, the total current flowing in the 
sample can be obtained by summing over the contributions due to the
electrons in the fully occupied extended states. 

Let us first compute the total amount of current carried in the edge
channel region. Following Trugman \cite{Trugman} we calculate the
Hall current carried by one LL
through some horizontal cross section $y = y_0$ of length $\triangle
L = x_2 -x_1$  along which the potential falls by $\triangle V = V_2 -
V_1$. The number of electrons per unit
area $n$ is equal to $ 1/2 \pi \hbar l_{b}^2 = eB/hc $. Define the
expectation
value of the local electron velocity operator at each point along the
line $ x_1 x_2$ \cite{Trugman}:

\be
   <\vec{v}(\vec{r})>=c \, \frac {\vec {\cal E}(\vec {r})\times \vec
    {B}}  {B^2},
\label{velocity}
\ee
where $ \vec {\cal E}(\vec {r})$ is the electric field. The current
$I_{x_1 x_2} $ that passes through the line $x_1 x_2$ is:

\be
    I_{x_1 x_2}= \int_{x_1}^{x_2}\! dx\, ne \, (\vec {v}\cdot \hat
    {y}) = -\frac {e^2}{h}\,\frac {V(x_2,y_0)-V(x_1,y_0)}{e}. 
\label{current}
\ee
Since local currents, circulating in the closed trajectories,
contribute neither to total diamagnetic current nor to non-equilibrium
Hall current (we will study their importance in Section \ref{sec:Streda}),
let us choose 
the points $(x_1,y_0)$ and $(x_2,y_0)$ on the extended
trajectories at potentials $V_1$ and $V_2$ respectively. Then the strip
between these two trajectories contributes the amount proportional to 
$\triangle V = V_2 - V_1$ to the total current as given by Eq.\
(\ref{current}).

The Hall conductivity is $\sigma_{xy}=I/V_{H}$, where $V_{H}$ is the Hall
voltage measured in the four-terminal resistance experiments. This voltage
is determined by the amount of current carried by each LL just
before the voltage contacts, as well as by the transmission and reflection
coefficients of these contacts. It is possible to show \cite{buttiker}
that the Hall conductivity will be quantized {\it only} if, upon leaving
the voltage contacts, all LL's are filled up to the same energy:
$\epsilon_{n,l}=\mu_{1}$, $\epsilon_{n,r}=\mu_{2}$ for all $n$. In
this case, the voltmeter measures the difference between the chemical
potentials at the two edges: $V_{H}=(\mu_2-\mu_1)/e$. Then from
Eq.\ (\ref{current}) follows that $\sigma_{xy}=Ne^{2}/h$. We must note
at this point that filling of each LL starts when the energy of this LL
drops to $\mu_{1}$ ($\mu_{2}$) at the left (right) edge.  This means
that {\it all} partially occupied states at one edge have the same energy
($\mu_{1}$ or $\mu_{2}$). Then, both in equilibrium {\em and} in the
non-equilibrium steady state described above, the total drop of the
potential $\triangle V_{n}$ in the part of the edge channel region
where the $n$-th LL is filled, equals to 
$(N-n)\hbar \omega_{c}$. Therefore, the potential falls by $(n-1)\hbar
\omega_{c}$ before the $n$-th LL starts to fill, and by $(N-1)\hbar
\omega_{c}$ through the whole edge channel region. As the energy difference
between the LL's in the compressible strips does not change when the
current is injected, the total potential drop $\triangle V_{n}$ is
also not changed by the injection of the Hall current. It
follows then from Eq.\ (\ref{current}) that the current flowing in the edge
channel region is always equal to the equilibrium diamagnetic
current, even when the non-equilibrium current is injected into the
sample. In other words, no non-equilibrium current flows in the edge
channel region, and the entire injected current is carried by the
states in the bulk of the system, to the right from the line $\Gamma$
on Fig.\ \ref{near_edg}a (see also Ref.\ \onlinecite{We_SSC}). 
The rest of the paper discusses the distribution of the extended
trajectories in the bulk of the sample. 

It is important to realize that the complete equilibration at the edge
leads, therefore, to complete absence of {\it non-equilibrium} current
in traditionally defined edge channels. On the other hand, if
this equilibration is not perfect, some portion of the 
{\it non-equilibrium} current is pushed into the edge
channels. However, in this case, the measured Hall conductance is not
quantized in units of $e^{2} /h$.
Experiments, involving non-ideal contacts \cite{alphenaar} (contacts
designed to avoid complete equilibration) clearly 
demonstrated this absence of quantization and implied presence of
currents in the edge channels. In fact, partial equilibration was
still achieved in such experiments due to inter-channel scattering. It
was argued, however, that the only source of perfect equilibration are
the {\it ideal} voltage contacts \cite{We_SSC}. We argue,
additionally, that employing the ideal contacts leads necessarily to
the absence of  {\it non-equilibrium} current in the edge
channels. Therefore, the  experimental conditions are crucial in
determining the Hall current distribution.


\subsection{The extended states, a continuity theorem}

Here we determine the spatial location of the extended states under
QHE conditions. Let us consider the behavior of the potential
immediately to the right of the line $\Gamma$. As mentioned above,
this potential initially falls due to the incompressibility in this
region (see Fig.\ \ref{near_edg}b). Exactly how far the potential
falls depends on the local configuration of the random potential, and
at some point the potential will necessarily start raising
again. Therefore, there is a spatially extended region to
the right of $\Gamma $ where the potential falls. Since $\Gamma$
itself is an equipotential line at energy $\epsilon_{\Gamma}= \mu - 
N  \hbar \omega_{c}$, by continuity there is a finite range of energies 
$\epsilon_{\rm last} < \epsilon  < \epsilon_{\Gamma} $ in which
equipotentials percolate in the $y$-direction. Here we denoted 
the lowest energy of these equipotentials by $\epsilon_{\rm last}$ and the 
chemical potential by $\mu$. 
We now ask what determines this minimum energy $\epsilon_{\rm last}$
and the shape of the corresponding equipotential. The results are far
from obvious.


\begin{figure}
  \begin{center}
    \leavevmode
    \epsfxsize=8cm
\epsfbox{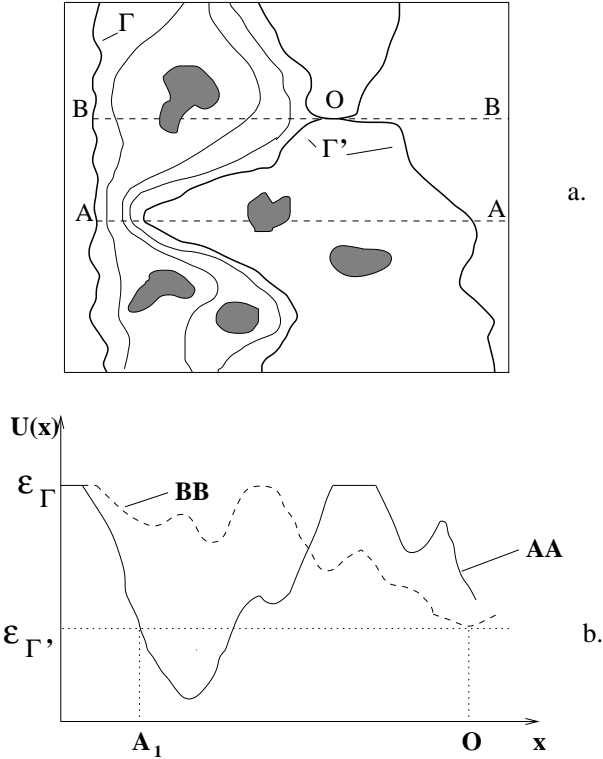}
  \end{center}
       \caption{Extended trajectories from
       the edge towards the bulk of the sample
       a. Open trajectories: $\Gamma$ is the inner boundary of the
        edge region, $\Gamma'$ is the last open trajectory in the bulk. 
        b. Corresponding energy profiles along the marked cross section
        lines AA and BB.
        {\label{bulk1}}}
\end{figure}


It was argued before that in the absence of the confining effects of
the physical boundary (we will call such an artificial construction a 
`quasi-infinite system'), the only effect of the finite size in
$x$-direction would be the existence of a narrow band of
equipotentials connecting left and right edges. 
Let us assume that an equipotential $\Gamma^{\prime}$ with energy
$\epsilon _{0}$ belongs to this band. 
It is useful to note that at this energy
there may also be other trajectories that do not percolate. 
Going back to a more realistic system, with confining potential at the
edges, one can see that this percolating trajectory $\Gamma^{\prime}$
is necessarily cut off by a rise of the potential towards the physical
edge (Fig.\ \ref{bulk1}). It is then easily
seen that at each energy between $\epsilon_{0}$ and
$\epsilon_{\Gamma}$ there exists one and only one equipotential line
open in $y$-direction,  and that no trajectory at energies lower than
$\epsilon_{0}$ can percolate vertically.

{\it Proof:} Let us connect lines $\Gamma$ and $\Gamma^{\prime}$
by means of an arbitrary line $\gamma$. Along this line the potential
has to fall from  $\epsilon_{\Gamma}$ at $\Gamma$ to $\epsilon_{0}$ at
$\Gamma^{\prime}$. The potential along the $\gamma$ line does not have
to be monotonous. 

\begin{figure}
  \begin{center}
    \leavevmode
    \epsfxsize=6cm\epsfbox{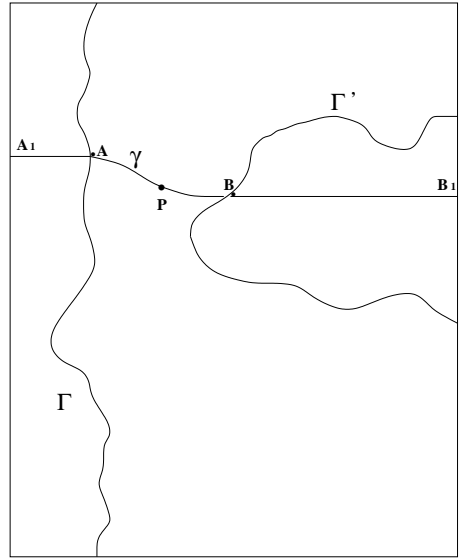}
  \end{center}
       \caption
      {Illustration of the proof of continuity theorem.
        {\label{proof}}}
\end{figure}

{\it a}.  Let us initially assume that the potential {\em is}
monotonous along $\gamma$. Now connect $\gamma$  with
the left and right edges by, say, horizontal straight lines $AA_{1}$
and $BB_{1}$ (Fig.\ \ref{proof}). Choose
any point $P$ on $\gamma$ between $A$ and $B$. We now ask whether an
equipotential line threading through $P$ percolates in the
$y$-direction. In  order for the trajectory passing through $P$ to be a
closed one, one has to find another point  on the line 
$A_{1}B_{1}$ with the same potential as in point $P$. 
Such point cannot be found on $\gamma$ because of the monotonous behavior
of the potential along it. There could be such points on the piece
$BB_{1}$ but in order to reach them one has to cross line
$\Gamma^{\prime}$ which is at different potential. Hence, there is no
way to close equipotential trajectory by going to the
right of $P$. A similar argument shows that one cannot close trajectory
to the left from $P$ either. Therefore, the equipotential passing through any
point on $\gamma$ is necessarily delocalized in the $y$-direction.

{\it b}. Going back to the general case of a non-monotonous potential
on $\gamma$, we argue that for any energy $\epsilon$ such that 
$\epsilon_{0} < \epsilon < \epsilon_{\Gamma}$ there will be an odd
number of points on $\gamma$: namely, there will be one more point on the
fall of potential than on the rise. Some of these points combine
pairwise into closed orbits. However, the additional single point
satisfies conditions similar to those stated in part {\it a}: the
equipotential passing through it must be extended in the $y$-direction.

From this discussion it is also clear that {\em no} trajectories with
energies beyond the interval $[\epsilon_{0}, \epsilon_{\Gamma} ]$ can
percolate in $y$-direction.


We immediately recognize that $\epsilon _{0}= \epsilon_{\rm last}$
and that the corresponding right-most vertically percolating 
equipotential is the line we denoted $\Gamma^{\prime}$ on Fig.\
3. Therefore, as a first approximation, the trajectory
$\Gamma^{\prime}$ has to percolate in both vertical and horizontal
directions. This is  a {\em critical trajectory} with at least one
saddle point $O$ on it. 
An equilibrium picture of extended trajectories emerges according to
this discussion as shown in Fig.\ \ref{bulk1}. 

Qualitatively, we conclude that at least the trajectories with
energies greater than, but sufficiently close to $\epsilon_{0}$, travel
throughout the system towards the `saddle point', as does the critical
trajectory itself. The latter comes near the boundaries only to get
around closed trajectories with similar energies and with 
large sizes in the $x$-direction. Since such trajectories are very
sparse, the critical trajectory spends most its path inside the 
system rather than near the boundary. In addition, there can also
exist vertically percolating trajectories in the edge compressible
regions to the left of line $\Gamma$ (see Fig.\ \ref{near_edg}).  


\subsection{General remarks}
\label{GR}

Insofar, by means of purely geometrical arguments, we have proved that:

{\bf i.} Contrary to a widespread belief, it is not just the existence
of the physical boundaries which is responsible for the formation of a
band of extended states (instead of a single extended state in the infinite 
random two-dimensional system). It is clear
from the preceding discussion that the number, density
and location of the trajectories extended in $y$-direction have
little to do with the sample size in $x$-direction. This is in
complete contrast with the common statement that the bandwidth of the 
extended states is inversely proportional to the sample size. The most
important factor is the {\it interaction between electrons} which
leads to the screening of the confining and disorder potential at the
physical boundary and, consequently, to the formation of an extended
boundary line $\Gamma$. It is the difference between the energy of
this equipotential line $\Gamma$ (which is equal to the chemical
potential $\mu$) and the energy of the state extended in a
corresponding infinite system, that determines the total number and
density of extended states in equilibrium.

{\bf ii.} In addition, we must conclude that even in equilibrium, 
extended states are present not only near the physical boundary but
also deep in parts of the bulk, their particular, 
sample-dependent distribution being determined by a specific
configuration of the disorder. 

In the next section we give a detailed description of how, under the
conditions of complete equilibration at each edge, the Hall current is
distributed in the sample.

\section{Delocalized trajectories and current distribution in the steady 
state}
\label{sec:steady}

Imagine going from the line $\Gamma$ towards the inside of the system
in the current carrying steady state. The difference from the
equilibrium state then is that the counterpart of $\Gamma$ at the
right edge is an equipotential at an energy lower than 
$\epsilon_{\Gamma}$ by the Hall voltage $V_H$. 
Using arguments similar to the ones applied in equilibrium, it is easy
to demonstrate that there is a critical trajectory $\Gamma^{\prime}$
with at least one saddle point in the steady state. The energy of this
critical trajectory will not be the same as in equilibrium: in our 
setting of Hall field directed from right to left, the following 
inequality should hold:  
{$(\epsilon_{\Gamma} - \epsilon_{\Gamma^{\prime}})_{\rm non-equil} >
(\epsilon_{\Gamma}-\epsilon_{\Gamma^{\prime}})_{\rm equil}$}.
Therefore, the number of extended trajectories differs from the 
one in equilibrium. This should be true independently of whether the
charge distribution is changed due to the presence of Hall
electric field or not.   

As the Hall current is injected into the sample, it charges the edge  
metallic regions establishing the chemical potential difference
between the two opposite edges. These charges produce the
long-range electric field decaying as the inverse
distance from each edge sufficiently far from this edge
\cite{Thouless,wexler,Thouless1,MacDonald}. 
In the context of the problem of Hall current distribution the effect
of charged edges was thoroughly studied in the
past~\cite{Thouless,wexler,Thouless1}. If these  
charges cannot move into the bulk they create a logarithmic
potential sufficiently far from the edges. This solution assumes zero
longitudinal conductivity. More realistically, however, one has to
estimate the charge relaxation times associated with the finite
longitudinal conductivity $\sigma_{xx}$ or with some other charge
relaxation mechanisms (not related to the transport by the extended
states) . Thouless \cite{Thouless1} considered a finite 
$\sigma_{xx}$, and assumed as the condition for a steady state a
constant value of dissipative current $j_{x}(x) = {\rm const}$. The
problem reduces to the two-dimensional Poisson equation and is solved
for the  charge distribution inside the Hall bar. This new charge
distribution produces a homogeneous electric 
field across the sample and can be
established only if there are regions in the bulk which can absorb
additional charges. The origin of such regions, their sizes and
spatial distribution will be discussed at the end of the paper in
the context of recent measurements, and for the estimates of typical
charge relaxation times. A more thorough treatment can be found in
other publications \cite{We_DOS,Dima-Th,We_Charging}. 

Consider first the situation when the relaxation 
time is very long compared with the time of measurement. 
In this case the Hall electric field is negligible at the saddle point on
the critical trajectory, falling off inversely with the distance from
the edge. One can argue that this critical
trajectory changes its energy but the saddle point on it does not
move. Consequently, all the {\it newly formed} percolating trajectories have
to be located between the critical trajectory and the adjoining edge
region, that is, in the region where delocalized states were present in
equilibrium. Similarly to equilibrium delocalized trajectories, the
newly formed ones are not distributed homogeneously. While in 
most of the area of the sample their average concentration follows the Hall
electric field (by the arguments used to prove the equilibrium
distribution), in the regions where the critical trajectory approaches
the edge, the density of newly formed delocalized states has to be
much higher. 

\begin{figure}
  \begin{center}
    \leavevmode
    \epsfxsize=6cm
    \epsfbox{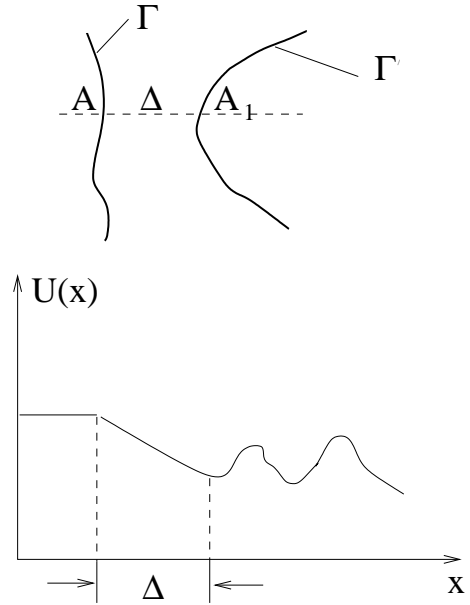}
  \end{center}
       \caption{Critical trajectory $\Gamma '$ approaching the
        boundary line $\Gamma$ and the form of electrostatic
        potential in this region (see the text).
                 {\label{Approach}}}
\end{figure}

Consider a horizontal cross section $AA_{1}$ where the 
critical line $\Gamma ^{\prime}$ approaches the line $\Gamma$ at a
distance $\Delta \ll W$ (Fig.\ 5). The fall of the potential
$\triangle V$  on the distance $\Delta $ due to the
external field is negligible compared to the Hall voltage. Yet,
the application of a small Hall field cannot destroy many localized
states inside the critical trajectory as in most parts of the system the
localizing disorder field is much greater than the external one. 
These two facts can be reconciled if
most of the newly formed delocalized trajectories with 
energies in the interval of the order of  $V_{H}$ appear in the narrow
neighborhood of the original position of $\Gamma^{\prime}$, to the
right of it. The trajectory 
$\Gamma^{\prime}$ itself shifts to the right by a distance $\delta$ to
let the vertically percolating trajectories penetrate into the interval
$AA_1$ with length $\delta$ (Fig.\ \ref{Approach}). Such a solution is
possible due to the fact that the slope of the potential in the
interval $AA_1$ is determined by the field
of disorder potential rather than by the much weaker external field.
In the areas of the sample where the critical trajectory is
sufficiently far from the edges so that the Hall field near this
critical trajectory is negligible,
the geometry of the critical line does not affect the position of the
field-induced delocalized states. The concentration of latter ones
follows the intensity of the Hall field and the concentration of
equilibrium extended trajectories: new trajectories are formed as a
response to the Hall electric field on those slopes of disorder
potential which contained delocalized states in equilibrium. One
concludes, therefore, that, except for the electrostatic Hall electric
field,  there are no specific features brought 
into the Hall current spatial distribution by the edges. Since 
equilibrium delocalized trajectories are generally {\bf not}
concentrated near the edge (previous Section), neither are the newly
formed ones. The density of non-equilibrium extended trajectories is
weighted towards the edges {\it only} to the extent the Hall electric
field is. 


The other limiting case is the condition of complete (fast) 
relaxation. It is realized if the charges, including those on
localized states, respond to the external electric field created by
the charged edges and redistribute themselves according to the steady
state solution. The typical rate of this response must be greater than
the rate at which the parameters are changed in the experiment.
The only possible source of charge redistribution are 
the metallic regions of partially filled Landau levels. 
Therefore, one has to estimate the time it takes for a finite
charge to be transfered from any such metallic region to the
neighboring one. 

There is experimental evidence that,
in many circumstances, equilibration does occur. Direct study of
relaxation processes \cite{Kukushkin} shows that the state of an
incompressible system achieved by raising the magnetic field from
zero to several Tesla at high temperature (of the order of 1K) differs
significantly from the one reached by increasing the magnetic field in
a pre-cooled (100mK) sample and then heating it up back to
1K. Although the relaxation considered 
in \cite{Kukushkin} requires spin-flip processes and is, therefore,
significantly slower, one can think of the relaxation as being
facilitated by higher temperatures allowing for greater hopping
or activated current. 

Other group of experiments produce some
indirect evidence of relaxation: the dependence of QH breakdown
current on the sample width. Most of these measurements show a 
linear dependence which contradicts the assumption that the current
flows primarily through the edges. This implies that 
that the electric field is close to a constant across the sample. This
can be achieved only by delivering 
electrons to localized states in the compressible islands. At currents
close to the critical current for breakdown of the QHE
the Hall electric field is comparable with the field of
disorder~\cite{We_BKDN} 
and this factor itself can initiate effective relaxation processes via
possible tunneling into unoccupied states on the higher Landau level
\cite{Heinonen}. However, right before the breakdown, the  bulk of the
system is a well-defined incompressible system, in which relaxation
processes are fast. We will present a consistent theoretical description
and quantitative conclusions in the next Section.

Consider now where the delocalized trajectories are 
located in such an equilibrated system in a weak Hall electric field. 
The picture of extended trajectories is changed relative to the
situation of an insulating (rigid) bulk only by the fact that the
electric field
is homogeneous on average rather than weighted towards the edges. This
difference, however, leads to the distribution of extended states being on
average a copy of such a distribution in equilibrium.
Therefore, the current injected into the system in such case flows in the
region between the lines $\Gamma$ and $\Gamma^{\prime}$ as they were
defined in equilibrium and has no preference of the edge regions. In
the other words, the current is {\it distributed} throughout the system.

\section{Implications of current distribution}
\label{sec:Streda}

In either of the two limits of short or long relaxation times, for
sufficiently small Hall currents, one can label the corresponding
regime as `linear response'. In such a `linear response' regime the
critical trajectory $\Gamma^ {\prime}$ plays the role attributed in
the edge-channels formalism to the innermost boundary of the edge
channels. However, unlike the edge channels, $\Gamma^{\prime}$ itself
is not confined to be near the physical boundaries. In this 
sense the description proposed here is complementary to the edge state
formalism. It also resolves some controversy around the validity of
St\v{r}eda formula \cite{Streda}.   

In the derivation of the St\v{r}eda formula it is assumed that the
derivative $ (\partial n / \partial B)_{\mu}$ is quantized due to
the existence of the gap in the density of states between Landau
levels. In this formula $ n$ stands for the density of states, $B$ for
magnetic field, and $\mu$ for the fixed chemical potential. This
assumption is crucial for the use of the Maxwell relation
$ (\partial n / \partial B)_{\mu} = (1/A)(\partial M / \partial
\mu)_{B}$, where $M$ is the magnetization and $A$ is the total area of
the system. Under certain assumptions (which we mention below) the
derivative in the right-hand side is equal to conductivity thus
providing the proof for the conductivity quantization. It is clear,
however, that there is just a {\it mobility} gap between LL's   
but there is a finite density of localized states in this mobility
gap. These localized states have to contribute to the derivative 
$ (\partial n / \partial B)_{\mu}$  in one way or another. This
immediately questions the validity of the assumptions used to prove
the  equality $\sigma_{xy}=(1/A)(\partial M / \partial \mu)_{B}$ ,
namely that the change in magnetization $\delta M$ is entirely due to 
additional edge current associated with the change in chemical
potential $\delta \mu$. 

For {\it interacting} electrons, however,  $\delta M$ is
constructed from both drift along extended equipotentials and
circulating currents on closed loops. The latter ones do not
contribute to the Hall conductivity but turn out to be essential in 
evaluating both right-- and left-hand sides of the Maxwell relation
\cite{Thouless2}. 
Our argument makes use of the existence of states in the mobility gap,
on the one hand by implying the formation of `new' extended
trajectories within this gap and on the other, by explaining how these new
extended trajectories become responsible for the total change in
magnetization. 

At higher Hall voltages one expects the original topology of states to
be gradually destroyed by the external field. The profile of the random 
potential has to change to absorb more delocalized trajectories
than it would be able to absorb in any of the two scenarios described
above. In fact, only at rather strong Hall currents does the treatment
presented in this Section need to be reconsidered. It happens only
when the relaxation processes and the dissipative transport change
significantly compared to the linear response regime. This, in turn,
occurs when the local Hall electric field, at least in some areas of
the sample, becomes comparable with the electric field of the disorder
potential. This field turns out to be of the order of $0.1
\hbar \omega_c / l_B$. The corresponding to Hall current densities are
$10$A/m which exceeds experimental values in most cases. At such
strong external fields the system becomes more homogeneous as the
potential fluctuations get wiped out. 
The newly formed percolating trajectories spread away from the
critical line $\Gamma^{\prime}$ and distribute throughout the sample.

\section{Interpretation of Recent Experiments}
\label{sec:Expts}

A theoretical model which captures most of the qualitative
and quantitative features of charge relaxation processes in the QH
regime has been constructed in  Refs.\ \onlinecite{We_Charging,Dima-Th}.  
Here we will sketch the main blocks
of the model and present the results leading to a consistent
explanation of the two recent experiments
\cite{Ashoori,MacEuen}. Here we concentrate on the experiments of
Ref. \onlinecite{MacEuen}.

The random resistor-capacitor \mbox{network} model (RRCNM) 
is based on the theory of charge transfer between 
isolated metallic islands in the bulk via hopping. 
Due to the small size of these islands one has to take into
account the Coulomb blockade effects: adding or removing a single
electron from the island costs a non-negligible charging
energy. Therefore, each island 
serves as a capacitor in the model. Typically the charging
energies are of the order of $1$ meV. Moving electron from a charged
island to an adjacent neutral one leads to either emission or
absorption of a phonon with the energy equal to the difference of the
charging energies of the two islands. This difference is 
typically $0.3 - 0.5$ meV, or $3-5$ K. 
Such phonon-assisted hopping is somewhat
suppressed, in a case of phonon absorption. However, in the range of
temperatures between $300$ mK and $3$ K in which most of the
experiments on current distribution and charge relaxation are
performed, this suppression is not dramatic.

The other factor limiting the rate of charge transfer through the bulk
is the distance between the communicating islands. The extent of the
wave function of an electron at the boundary of a compressible island
is of the order of a few magnetic lengths. If the distance between the
islands significantly exceeds this width, the overlap between the
states participating in an individual act of hopping is exponentially
small leading to a vanishing hopping probability. This limitation
defines a few (if any) paths which are available for a charge
transfer. Note that this process not only leads to a finite
longitudinal current but also to the formation of a new steady state
with a charge distribution different from the one in equilibrium.
The links between the islands enter into the model as resistive
elements. 

The average distance between adjacent islands varies substantially
with the filling factor. In the range of temperatures mentioned above,
the QH plateau is quite narrow: it spans just $5\%$ to $20\%$ of the
entire distance between the Landau levels on the scale of filling
factors. The greater the number of filled LL's, the narrower the
plateau and the smaller the typical distance between the compressible
islands. It turns out that relaxation can be very slow for the filling
factor $\nu =2$ (spin-unresolved LL's)  at $T \simeq 300$ mK. However,
for $\nu = 4$, even at these low temperatures the RRCNM predicts
relaxation times of the order of $1$ ms which is shorter than the
typical experimental time scales. 

\begin{figure}
  \begin{center}
    \leavevmode
    \epsfxsize=6cm
    \epsfbox{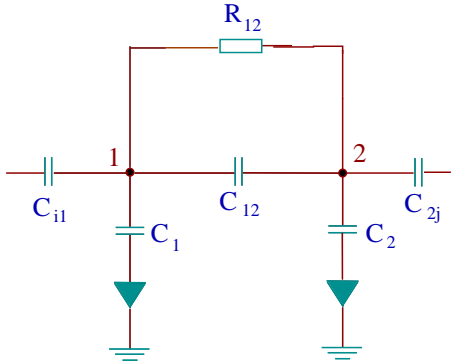}
  \end{center}
       \caption[The Random Resistor-Capacitor Network]
                {The basic block of the random resistor-capacitor
                 network.
                 {\label{RRCN}}}
\end{figure}

The basic block of the simplest RRCN is shown on Fig.\ \ref{RRCN}. Each
node corresponds to a compressible island which is connected
resistively to a neighboring one. The capacitance of each island to
the ground is accounted for by the capacitors $C_i$ while mutual
capacitances between the islands $i$ and $j$ are denoted as $C_{ij}$. 


The RRCNM is solved numerically for each particular configuration of
disorder and at different filling factors. The solution provides the
geometric paths of electron transfer, the charges on compressible
islands, AC and DC responses of the system, and charge relaxation time
$\tau$. Given the conditions of an experiment, one can obtain these
characteristics to predict or explain the observed results. 

\begin{figure}
  \begin{center}
    \leavevmode
    \epsfxsize=5cm
    \epsfbox{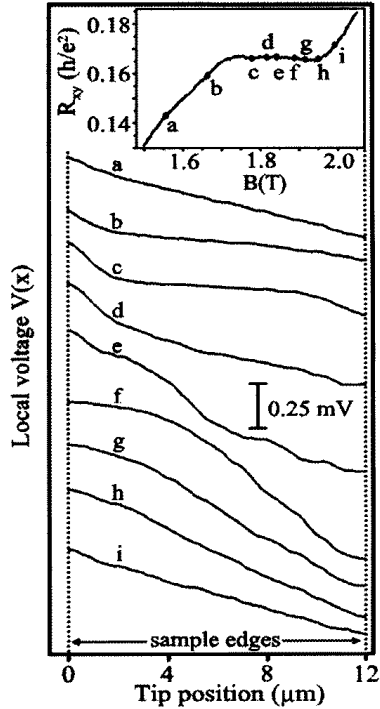}
  \end{center}
       \caption[Voltage profiles in the QH bar near $\nu=6$. After
        McEuen {\it et al}]
                {Voltage profiles across a 12 $\mu$m-wide Hall bar on a
low-mobility sample near the $\nu=6$ QH plateau (traces offset for
clarity). The B-fields at which the profiles were measured, and the
$\nu=6$ plateau measured by transport are shown in the inset.
(a) Well below the plateau, the Hall voltage profile is linear. (b,c)
As the plateau is approached, the bulk retains a linear profile but
decouples from the edges. (d-g) Well inside the plateau, the profile
in the bulk becomes complicated and changes rapidly with field. (h,i)
A linear profile returns upon leaving the plateau. [after Ref.\
\onlinecite{MacEuen}] 
                 {\label{AFM}}}
\end{figure}

In recent measurements \cite{MacEuen}, the atomic force microscope
technique has been used to study the distribution of the Hall
current in the QH sample. Their results which we interpret below, are
shown on Fig.\ \ref{AFM}.


The scheme of this experiment is different from the one discussed in
section \ref{subsec:start}. This is {\it not} a transport-type
measurement which means that there are no voltage contacts attached to
the sample. The voltage profile is measured capacitively, without a 
direct contact with the 2DEG. Therefore, the arguments of the section
\ref{subsec:start} proving the absence of non-equilibrium Hall current
in the edge channels, will not apply to this experiment in
general. More specifically, there is no ideal equilibration between
the edge channels provided by the voltage contacts in transport
measurements. The 
only source of inter-edge-channel equilibration is the mutual
scattering of electrons \cite{Chklovskii}. It is less effective than
the equilibration in the voltage contacts but for sufficiently narrow
edge channels it still leads to the establishment of equal chemical
potentials in the compressible edge regions of different LL's.

On Fig.\ \ref{AFM} the results of Hall resistance $R_{xy}$
measurements are plotted. The curve \ref{AFM}(a) corresponds to a
filling factor $\nu \simeq 6.8$. The bulk is clearly a metallic
system. There are three (spin-degenerate) compressible edge channels
($0\!<\nu<\!2$, $2\!<\nu<4$, and $4\!<\nu\!<6$) and three
incompressible edge channels ($\nu \! = 2, 4, 6$). As shown in
Ref.\ \onlinecite{Chklovskii}, only the inner-most edge regions can be
wide -- up to $10 l_B$. Since the resolution of the experiment is
about $20 l_B$, 
the edge regions, whether they are equilibrated or not, cannot be
resolved. This leads to a linear profile of the Hall voltage.

\begin{figure}
  \begin{center}
    \leavevmode
    \epsfxsize=7cm
    \epsfbox{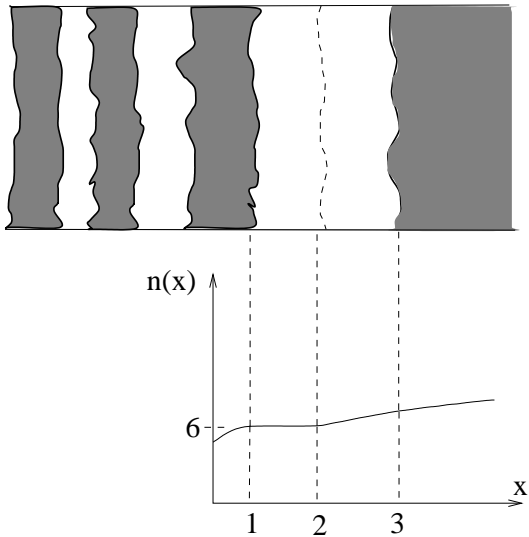}
  \end{center}
       \caption[Interpretation of the experiment by McEuen {\it et al}]
{Interpretation of the experiment in Fig.\ \ref{AFM} (Ref.\
\onlinecite{MacEuen}). The  
structure of the 2DEG near the edge. Top: alternating compressible (shaded)
and incompressible (blank) regions. Bottom: filling factor
profile. The dashed line on the top picture and point 2 on the bottom
correspond to the distance where the top-most 8-th LL starts filling
according to the electrostatic solution \cite{Chklovskii}. In reality,
the preceding incompressible edge channel with $\nu=6$ extends
deeper into the bulk (point 3) due to the presence of disorder (see
the text). 
                 {\label{Interpret}}}
\end{figure}

As the QH plateau $\nu = 6$ is approached with the increasing magnetic
field, the bulk filling factor decreases. On the curve
\ref{AFM}(b) the bulk filling factor is $\nu_0 \simeq 6.2$. The
filling profile is sketched on Fig.\ \ref{Interpret}. One can use the
electrostatic solution for the 2DEG density at the edge,
Eq.\ (\ref{Chk_Den}), to estimate how fast the top-most 7-th LL gets
filled to its bulk value. Assuming for the depletion width $l$ the
typical value  $l \approx 10 l_B$, we find that, for $\nu_0 = 6.2$
the distance between the line with $\nu=6.0$ (point 2 on
Fig.\ \ref{Interpret}) and $\nu = 6.1$ (point 3 on the same picture) is
approximately $300 l_B$ compared to just $10 l_B $ for $\nu_0 =
6.8$. Both experimental evidence of the final widths of QH plateaus and
our numerical simulations
\cite{We_DOS} clearly demonstrate that the system remains incompressible
in a certain range of filling factors $\triangle \nu$ around the integer.
Given the parameters used in these experiments
(mobility $80 000$ cm$^2$/Vs, density $2.8$ x$10^{11}$ cm$^{-2}$ and
temperatures between 0.7-1.0 K), according to our numerical
simulations (details of which can be found in Ref. \onlinecite{We_DOS}), 
$\triangle \nu \ge 0.1$. Therefore, on a length of the
order of $300 l_B$ to the right from point 2 on Fig.\ \ref{Interpret}
the system is still {\it incompressible}. Such a wide incompressible
strip makes mutual equilibration between the inner-most compressible
edge channel ($4 < \nu < 6$) and the outer-most states of the
compressible bulk practically impossible. To find out where
the Hall voltage accumulates one has to study local longitudinal
conductivity $\sigma_{xx}$. Since the bulk is metallic, its $\sigma_{xx}$
is quite large (typical metallic conductivity is of the order of $0.1
e^2/h$).
On the other hand, the longitudinal conductance of a wide
incompressible strip is exponentially small. The Hall current flows in
the region with smaller dissipative conductivity. This explains the
observation on Fig.\ \ref{AFM}(b),(c) of large potential drops near the
edges. 


As the bulk filling factor approaches $\nu_0 = 6$ even closer, the bulk
becomes less and less conductive and the incompressible edge region
gets wider (eventually connecting the two edges and transforming the bulk
into the incompressible state at $\nu_0 \approx 6.1$). This leads to a
smooth transition from an almost linear potential profile across the
entire sample Fig.\ \ref{AFM}(d),(e) to a regime where the potential
is flat at the edges and falls only in the bulk (Fig.\
\ref{AFM}(f),(g)). The profile of the potential in the last regime
confirms two conclusions we have made earlier in this paper. The
linear fall of the potential in the bulk means there is a complete
charge relaxation. In fact, the numerical 
calculations based  on the RRCNM predict relaxation times of the
order of $10^{-5}$s for the experimental conditions. This is by far
the shortest time present in the problem. Second, oscillations of the
curves around the linear profile are caused by the inhomogeneities in
the bulk: a linear average electrostatic potential is produced by
the charged metallic islands which vary in size and in separation from
each other.

\section{Conclusions}

We have proved, by means of purely geometrical arguments, that a band
of extended states, extended from source to drain, results from the
interplay between electron interactions and random and confining
potentials. These delocalized states extend through the bulk of the
system (see Sec.\ \ref{GR}).

Based on these conclusions we built a consistent
description of the delocalized states in the regime when classical
percolation theory is applicable. 
We were able to demonstrate that, under the most
general conditions, the Hall current-carrying states are distributed
in the bulk of the Hall sample rather than confined to the edge
region. We showed how particular experimental conditions 
influence the results of the measurements, and yet confirm the general
picture of distributed Hall currents. 
We believe that the controversy regarding the Hall current
distribution, as well as the nature and location of extended states in
the quantum Hall system is now resolved. 

The low temperature regime, however, still needs to be
addressed since the equivalence between equipotentials and electron
trajectories is lost when quantum effects become essential (for an
extended discussion on the transition between the two regimes see
Refs. \onlinecite{We_DOS,Kir_Thesis}). The authors do not know of any
experiment on the distribution of the Hall current and electric fields
which explores this low temperature regime. Theoretically a
more complete understanding of the microscopic structure of disordered
quantum Hall liquid is necessary to solve this problem.

\acknowledgements

The authors are grateful to Prof.\ David Thouless for his enlightening
comments and attention to this work. We thank Jung Hoon Han for
numerous helpful discussions throughout this work, and  B. Altshuler
and A. L. Efros for useful discussions at the early stages of
development. Finally, we thank the group of P. McEuen, R. Ashoori,
S. Tessmer, and A. Yacobi for sharing their preliminary experimental
results  prior to publication. This work was supported by the NSF,
Grant No. DMR-9528345. CW was supported in part by Grant No. DMR-9628926.




\end{document}